\documentclass[twocolumn,showpacs,superscriptaddress,preprintnumbers,amsmath,amssymb,epsfig,floatfix,aps,bibnotes]{revtex4-1}\usepackage{amssymb}

\usepackage{lipsum}
\usepackage{amsmath}
\usepackage{upgreek}
\usepackage{epsfig}
\usepackage{graphicx}
\usepackage{dcolumn}
\usepackage{color}
\usepackage{natbib}  
\usepackage{hyperref}
\usepackage{breakurl}
\usepackage{mathrsfs}
\hypersetup{colorlinks=true, citecolor=blue, urlcolor=blue, linkcolor=blue}
\usepackage{bm}
\usepackage{tabularx}
\newcolumntype{L}[1]{>{\raggedright\arraybackslash}p{#1}}
\newcolumntype{C}[1]{>{\centering\arraybackslash}p{#1}}
\newcolumntype{R}[1]{>{\raggedleft\arraybackslash}p{#1}}

\usepackage[capitalize]{cleveref}

\begin{document}
\title{Gate-dependent vacancy diffusion in graphene}
\author{Rohit Babar}
\affiliation{Department of Physics, Indian Institute of Science Education and Research, Pune 411008, India}	
\author{Mukul Kabir}
\email{Corresponding author: mukul.kabir@iiserpune.ac.in} 
\affiliation{Department of Physics, Indian Institute of Science Education and Research, Pune 411008, India}
\affiliation{Centre for Energy Science, Indian Institute of Science Education and Research, Pune 411008, India}
\date{\today}

\begin{abstract} 
Kinetics of vacancy defect in graphene drives structural modifications leading to disorder, multi-vacancy complex and edge reconstruction. Within the first-principles calculations, we study the dynamic Jahn-Teller distortion and diffusion of a vacancy defect. Further, the intricate dependence of carrier doping is systematically investigated. The experimental observation of dynamic Jahn-Teller distortion is argued to be blocked by defect functionalization and charge doping. We demonstrate that lattice relaxation perpendicular to the graphene sheet along with the in-plane strain relaxation play predominant roles in predicting the correct microscopic mechanism for vacancy diffusion. The importance of quantum correction to the classical barrier is discussed. The calculated activation barrier increases upon both electron and hole doping and the observed trends are explained by the differential charge density distribution and hardening of the responsible low-energy phonon modes. Electron doping essentially freezes the vacancy motion, and thus any degradation mediated by it. While tracking and analyzing the vacancy diffusion experimentally in graphene is a difficult task, the present results will motivate new experimental efforts and assist interpretation of the results. 
\end{abstract}
\maketitle

\section {Introduction}
Schottky lattice defects can significantly alter the physical, chemical and magnetic properties of graphene.~\citep{10.1021/acsnano.5b01762,10.1021/nn102598m,10.1126/science.1142882, PhysRevLett.102.236805,10.1038/nmat2830} Lattice vacancies are thermodynamically created at a finite temperature, originate from non-equilibrium growth such as micro-mechanical cleavage, chemical vapor deposition, and growth on a substrate; and could also be generated by electron and ion irradiations.~\citep{10.1126/science.1142882,10.1038/nature02817, 10.1021/nl8010337, 10.1021/nl801386m, PhysRevB.78.233407, 10.1063/1.3062851, PhysRevLett.104.096804, 10.1038/nature09718, 10.1021/ja403583s} The single vacancy defect is the most important point defect,~\citep{10.1021/nn102598m} and has been identified using aberration-corrected high-resolution transmission electron microscopy,~\citep{10.1038/nature02817,10.1038/nature05545,PhysRevLett.106.105505,10.1021/nn401113r, C4NR01918K} and scanning tunneling microscopy.~\citep{PhysRevLett.104.096804} While a single carbon atom is removed from the hexagonal graphene lattice, three dangling bonds are created on the adjacent atoms. This symmetric defect structure concurrently undergoes a spontaneous Jahn-Teller distortion to a symmetry-reduced 5$-$9 ring, where two of the neighboring atoms bond weakly leaving one unsaturated dangling bond.~\cite{10.1021/nl801386m} Further, the Jahn-Teller distorted vacancy structure can easily switch between three equivalent orientations with a very small kinetic barrier,~\citep{PhysRevB.68.144107} however, which is debated experimentally.~\citep{10.1021/nn401113r} With a $\sigma$ contribution from the dangling bond, the $\pi$ electron imbalance generates a semi-localized magnetic moment at the defect site according to the Lieb's theorem for bipartite hexagonal lattice.~\citep{PhysRevB.75.125408,10.1038/nphys2183, PhysRevB.85.245443,10.1038/ncomms3010} 
This localized moment also observed to undergo many-body Kondo screening due to the interaction with conducting electrons.~\citep{10.1038/nphys1962}

The interaction between spatially scattered vacancy defects and their kinetics can drive structural modifications including the formation of disordered regions.~\citep{10.1021/nn102598m} Migration of mono-vacancies and their eventual coalescence lead to multi-vacancy pores.~\citep{10.1038/nnano.2009.194, Lehtinen2013} A directional strain field can lead to the formation of line defects due to multiple vacancy mergers.~\citep{PhysRevB.78.165403} The biased vacancy migration toward the edge of a graphene flake alters the edge character at room temperature and thus severely alter the electronic, magnetic and transport properties.~\citep{Girit1705,Santana201380,Du20151270}  These defects act as strong scattering centers and disrupt the ballistic nature of electronic transport in graphene, and thus crucial to device performance.~\citep{10.1021/acsnano.5b01762,10.1021/nn102598m} Thus, the single vacancy diffusion acts as a microscopic unit process for the formation of higher order defect complexes. In this context, a profound understanding of the microscopic mechanism for vacancy propagation is necessary, especially at the device operating conditions. 


While the vacancy diffusion in three-dimensional crystals is being studied for decades, the same for the two-dimensional materials are exceedingly difficult to track and interpret. The presence of a substrate, layer thickness, external perturbations, as well as the interaction with the experimental tracking device such as STM may play a significant role and the interpretation for the results become that much difficult. Thus, the experimental data on vacancy migration and simultaneous estimation of the barrier in graphene is absent.  In contrast, vacancy diffusion on the graphite surface was studied by scanning tunneling microscopy.  Using the measured vacancy jump frequency as a function of temperature, the activation energy of 0.9$-$1.0 eV was estimated assuming a pre-exponent factor of 10$^{13}$ s$^{-1}$.~\citep{10.1021/jp901578c} In the present context of vacancy diffusion on the single-layer graphene lattice, this estimated barrier on the basal plane of graphite should be treated as the upper bound. 
In contrast, the vacancy defect has been studied within the first-principles calculations, though the microscopic diffusion mechanism is still debated.  The theoretical activation energy for diffusion to the nearest lattice sites are estimated between 1.1$-$1.4 eV range.~\citep{PhysRevLett.95.205501, 10.1016/j.cplett.2005.10.106, 10.1016/j.diamond.2010.06.010, C3NR06222H, 0953-8984-26-11-115303, 10.1016/j.cplett.2016.02.005} The large inconsistency between the experimental activation barrier estimated from the diffusion on the graphitic surface and the theoretical prediction on the single-layer graphene may result due to inappropriate consideration of strain relaxation during the vacancy migration, which eventually leads to the inaccurate transition-state.

Thus, thermally activated migration of point-defects causing degradation needs to be better understood, predicted and controlled. Here, we report a comprehensive study on the microscopic mechanism of vacancy diffusion. We elaborate the importance of strain relaxation during diffusion and the effects of the gate voltage in a device setup are investigated.
We also investigate the dynamic Jahn-Teller distortion and discuss the impossibility of experimental observation due to defect functionalization and charge doping.  The effect of applied gate voltage is studied through varied carrier doping within the experimentally realized carrier concentrations of 5$\times$10$^{13}$ cm$^{-2}$.~\citep{10.1038/nnano.2008.67} We illustrate that both out-of-plane and in-plane strain relaxations are essential to predict the accurate mechanism and the corresponding activation barrier.  The counterintuitive dependence of activation barrier on the gate voltage is explained through the low-energy phonon modes and differential charge density distribution. The quantum Wigner correction to the classical activation barrier at finite temperature is discussed.  The present results suggest that vacancy migration will considerably slow down under both positive and negative gate voltage, and thus will decelerate the concurrent graphene degradation in a device setup.


\section{Computational Details}
Calculations were carried out using the spin-polarized density functional theory (DFT) as implemented in the Vienna {\em ab initio} simulation package.~\citep{PhysRevB.47.558,PhysRevB.54.11169} The ion core and the valence electrons were described within the projector augmented wave formalism,~\citep{PhysRevB.50.17953} and the wave-functions were expanded in the plane-wave basis with 500 eV cutoff for the kinetic energy.  The exchange-correlation energy was computed using the Perdew-Burke-Ernzerhof (PBE) form of generalized gradient approximation (GGA).~\citep{PhysRevLett.77.3865} All the structures were allowed to fully relax until all the force components were less than 0.01 eV/\AA\ threshold, where the Brillouin zone was sampled using a 2$\times$2$\times$1 Monkhorst-Pack $k$-grid.~\citep{PhysRevB.13.5188} A finer 8$\times$8$\times$1 $k$-grid was used to calculate the density of states (DOS). Calculations were carried out with a 10$\times$5 supercell repeated with the rectangular $(\sqrt{3},3)a_0$ unit-cell, where $a_0$ is the nearest-neighbor distance between carbon atoms. Thus, this supercell consisted 200 carbon atoms without the vacancy, and all the results represent this supercell if not otherwise stated. We have also studied and analyzed our results with a smaller 6$\times$4 $(\sqrt{3},3)a_0$ and a larger 13$\times$8 $(\sqrt{3},3)a_0$ rectangular supercells consisting 96 and 416 atoms, respectively. A larger 4$\times$4$\times$1 Monkhorst-Pack $k$-grid was used for the smaller smaller 6$\times$4 $(\sqrt{3},3)a_0$ supercell. The periodic images perpendicular to the graphene sheet was separated by  12 \AA\ vacuum to cancel any spurious interactions. The phonon frequencies were calculated at the $\Gamma$-point for selected atoms around the vacancy using the density functional perturbation theory (DFPT). The effect of external gate voltage was simulated by varying the carrier (electron and hole) density. The microscopic mechanism for vacancy diffusion and the corresponding activation energy was determined using the climbing image nudged elastic band method (CI-NEB).~\citep{1.1329672} The minimum energy path was confirmed by the presence of a single imaginary frequency in the vibrational spectra for the transition state configuration.


\section{Results and Discussion}
We start our discussion with the thermodynamic stability of different vacancy structures and investigate the associated strain field. The Jahn-Teller (JT) distorted planer vacancy with 5-9 ring structure V$_1(5|9)$  and one unsaturated dangling bond is found to be the ground state with 7.65 eV formation energy, which is in agreement with the previous theoretical and experimental results.~\citep{PhysRevB.68.144107,PhysRevB.93.165403,10.1021/nn401113r}  The other JT distorted vacancy structure V$_1'$ is found to be metastable with 0.2 eV higher energy. Further details are in Supplemental Material.~\citep{supple}

\begin{figure}[t!]
 \begin{center}
\rotatebox{0}{\includegraphics[width=0.30\textwidth]{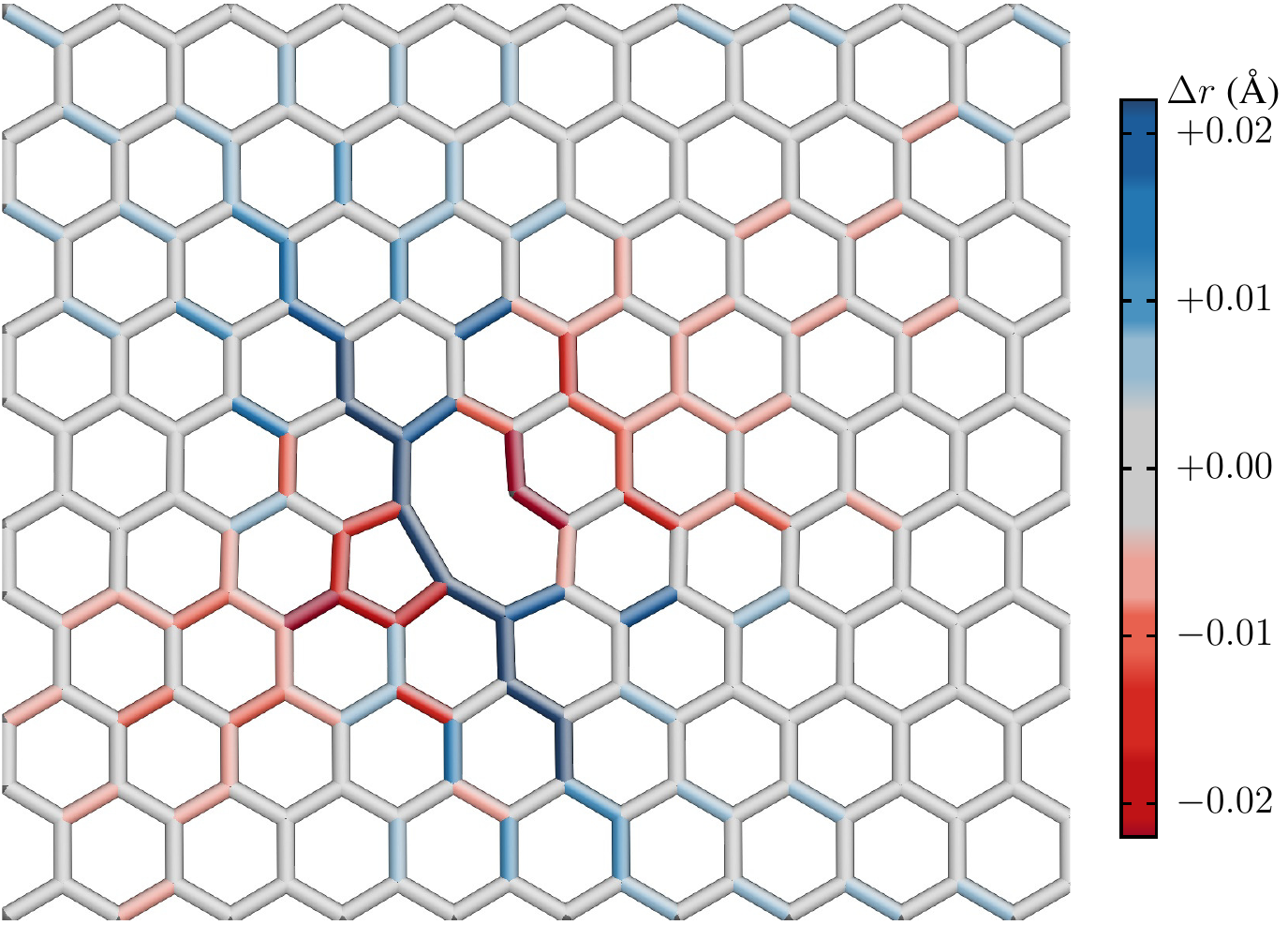}}
 \caption{Defect induced strain field for V$_1(5|9)$ calculated as the difference in bond length $\Delta r$ with defect free graphene lattice. Elongation and contraction of bonds are represented in blue and red colors, respectively. A significant strain field is localized in the proximity of the defect, while the amplitude decreases with distance from the defect and extends over 2 nm.}
 \label{fig:figure1}
 \end{center}
 \end{figure}

In the context of vacancy diffusion, it is anticipated that the strain field generated in the graphene lattice due to the point-defect should play a critical role. Such strain field is predicted to be long-range in graphene and other carbon nanostructures.~\citep{Santana201380,s00214-011-0910-3,PhysRevLett.109.265502, PhysRevB.89.201406, j.carbon.2016.02.064} Further, complex defects such as dislocation in graphene generates substantial corrugation in addition to in-plane strain extended over several nanometers.~\citep{Lehtinen2013,PhysRevB.80.033407,warner2012dislocation} Moreover, it has been demonstrated earlier that defect-defect interaction could severely alter mechanical, electronic and physical properties of graphene.~\citep{j.carbon.2016.02.064,PhysRevB.86.075402} Thus, to study an individual vacancy, the defect-defect interaction which originates through long-range strain field must have an inappreciable effect. 

We calculate the strain field produced in the lattice due to the V$_1(5|9)$ defect by calculating the difference in C$-$C bond distances in defect-free graphene without the defect (Fig.~\ref{fig:figure1} and Supplemental Material~\citep{supple}). The strain field in the vicinity of the point defect is evident and is extended over 2 nm as the amplitude decreases with the distance from the defect. Thus, a supercell extending over 2 nm is necessary to study an isolated point defect, and a rectangular 10$\times$5 $(\sqrt{3},3)a_0$  supercell is found to be sufficient. In this regard, a comparatively smaller 6$\times$4 supercell is found to be insufficient due to finite strain field at the cell boundary, which will influence the isolated defect properties through interaction with the periodic image. In contrast, a larger 13$\times$8$ (\sqrt{3},3)a_0$ supercell is found to be redundant, which does not show any significant difference in the strain field while compared with the 10$\times$5 $(\sqrt{3},3)a_0$ supercell.~\citep{supple}  In contrast to V$_1(5|9)$ defect, the strain field generated at the metastable V$_1'$ defect configuration is much localized.~\citep{supple} Such different strain fields generated by the different types of defects are in good agreement with previous experimental observations.~\citep{10.1021/nn401113r}

\subsection{Rotational reconstruction dynamics of V$_1(5|9)$}
There are three equivalent 120$^\circ$ rotated degenerate configurations for the V$_1(5|9)$ defect, which can be accessed by the dynamic JT reconstruction. The local stretching and concurrent reformation of a new pentagonal bond lead to these degenerate configurations. Note that such reconstruction does not involve any mass diffusion, and the energy requirement for this bond reorientation mechanism is expected to be small. The barrier for this local swapping of the reconstructed bond is calculated to be 0.2 eV, with V$_1(5|9)$ $\rightarrow$ V$_1'$  $\rightarrow$ 120$^\circ$-V$_1(5|9)$ as the mechanism. However, a wide range of barriers 0.13 -- 0.78 eV have been reported earlier.~\citep{0953-8984-26-11-115303, 10.1016/j.cplett.2016.02.005,PhysRevB.68.144107,10.1021/jp512886t} The small energy cost indicates the  V$_1(5|9)$ defect should undergo a  continuous reconstruction at a moderate temperature. The cycle frequency for the dynamic reconstruction can be estimated $k \sim k_0\exp(-\Delta E/k_BT)$, where $k_0 \sim 10^{13}$ Hz is the attempt frequency. Thus, the cycle frequency at 300 K is $\sim$ 4 GHz, which considerably slows down at low temperature to $\sim$ 1 Hz at 77 K. The corresponding time-scale is thus about 230 ps and 1 s, respectively, at these temperatures. However, these time-scales are much shorter compared to the experimental observations, which do not witness reconstruction for 150 s,~\citep{10.1021/nn401113r} and would be interesting to address the discrepancy. One possibility could be that functionalization of the under-coordinated C-atom blocks the dynamic reconstruction. The other possibility is the charge transfer between the dielectric TEM grid and graphene, which could essentially hinder the dynamic reconstruction of the V$_1(5|9)$ defect. It is known that charge is transferred to graphene from the dielectric Si$_3$N$_4$ TEM grid as is used in the experiments.~\citep{10.1021/nn401113r,10.1063/1.3623567}

\begin{figure}[t!]
 \begin{center}
\rotatebox{0}{\includegraphics[width=0.47\textwidth]{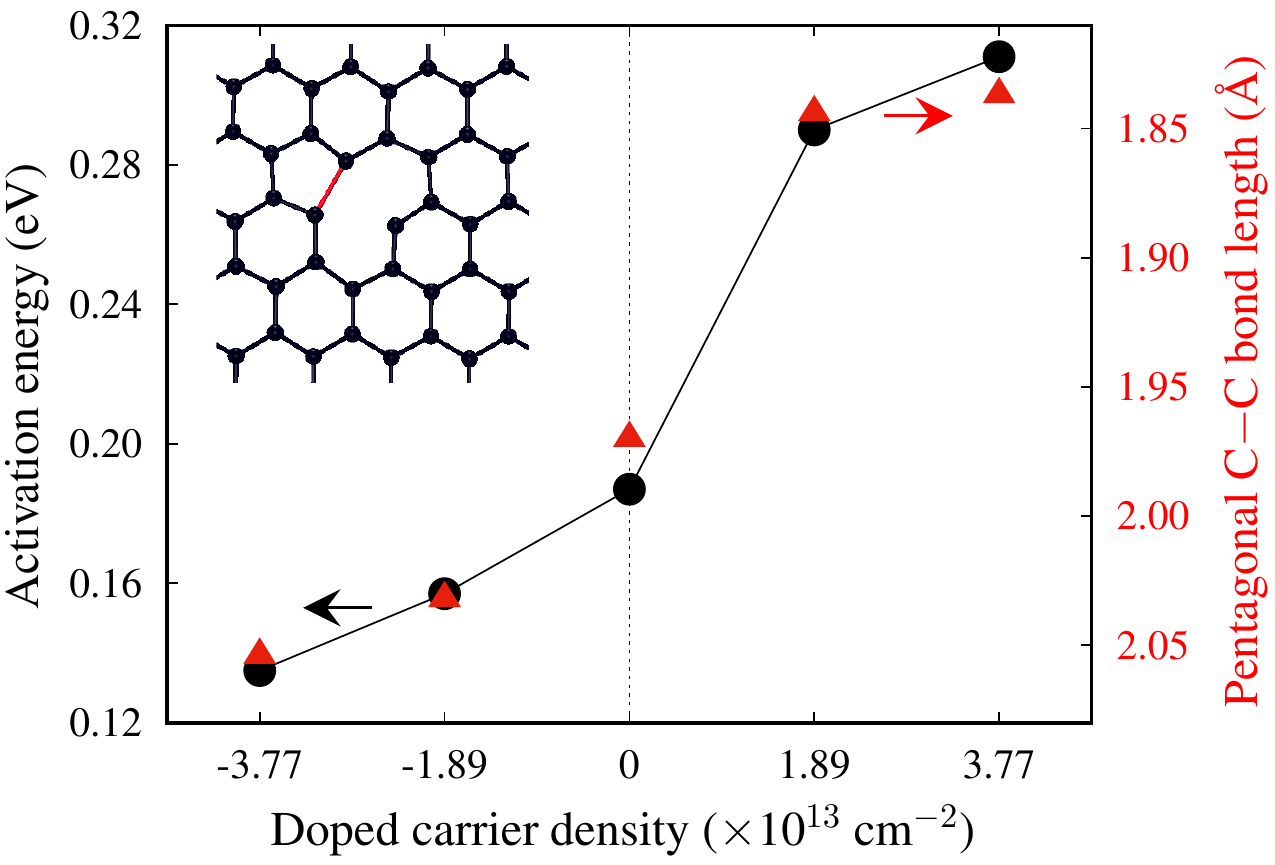}}
 \caption{The activation energy for the dynamic JT reconstruction, V$_1(5|9)$ $\rightarrow$ V$_1'$  $\rightarrow$ 120$^\circ$-V$_1(5|9)$, is strongly affected by carrier doping in the graphene lattice. The overall trend can be explained by the corresponding change in the pentagonal C$-$C bond (shown in inset) due to charge doping leading to strengthening or weakening of C$-$C bond upon electron and hole doping, respectively.}
 \label{fig:figure2}
 \end{center}
 \end{figure}

Once a carbon atom is removed from the graphene lattice, the time-scale for $D_{3h}$ to JT reconstruction is much shorter than the typical H diffusion on the graphene sheet.~\citep{PhysRevB.84.075486, PhysRevB.80.085428} Thus, the V$_1(5|9)$+H is the ubiquitous defect complex and survives at high temperatures even beyond 600 K.~\citep{PhysRevB.89.155405, CASARTELLI2014165} The earlier muon spin-resonance spectroscopy also indicates to singly hydrogenated vacancies.~\citep{10.1021/nl202866q} We, therefore, investigate the dynamic JT reconstruction in the presence of H. The H functionalization decreases the pentagonal C--C bond to 1.92 \AA, and strongly influences the breaking and subsequent reformation of this bond.  Therefore, the calculated activation energy for the dynamic JT distortion for the hydrogenated vacancy, V$_1(5|9)$+H  $\rightarrow$ 120$^\circ$-V$_1(5|9)$+H, is calculated to be much larger to 1.28 eV (Supplemental Information~\citep{supple}).

Now we study the effect of carrier doping on the dynamic JT reconstruction.  With the addition of electrons (holes) to the graphene lattice, the energy requirement for the dynamic JT reconstruction increases (decreases) as shown in Fig.~\ref{fig:figure2}. The doped carrier is semilocalized at the defect site as observed from the differential charge density (will be discussed later), and strongly affects the pentagonal C$-$C bond. The overall trend in activation barrier can be explained by the change in pentagonal C$-$C bond in V$_1(5|9)$, which is reduced under electron doping, and conversely increased while doped with holes (Fig.~\ref{fig:figure2}). Thus, compared to the neutral V$_1(5|9)$ defect, strengthening (weakening) of the pentagonal C$-$C bond due to electron (hole) doping results in higher (lower) energy requirement for the local swapping of the reconstructed bond. The increased energy requirement due to H functionalization and electron doping has significant implications on the impossibility of its experimental observation till date. With 1.28 eV activation barrier, the H functionalization essentially restricts the dynamic JT reconstruction process. In contrast, the energy cost for local reconstruction increases to 0.29 eV for 1.89$\times$10$^{13}$ cm$^{-2}$ electron doping with a reconstruction time-scale of about 10$^6$ s at 77 K. Therefore, even though the dynamic JT reconstruction is possible, the vacancy functionalization or the charge doping in experimental situations may essentially lock the V$_1(5|9)$ defect structure, and hinder experimental observation of dynamic reconstruction.

\begin{figure}[t!]
 \begin{center}
\rotatebox{00}{\includegraphics[width=0.47\textwidth]{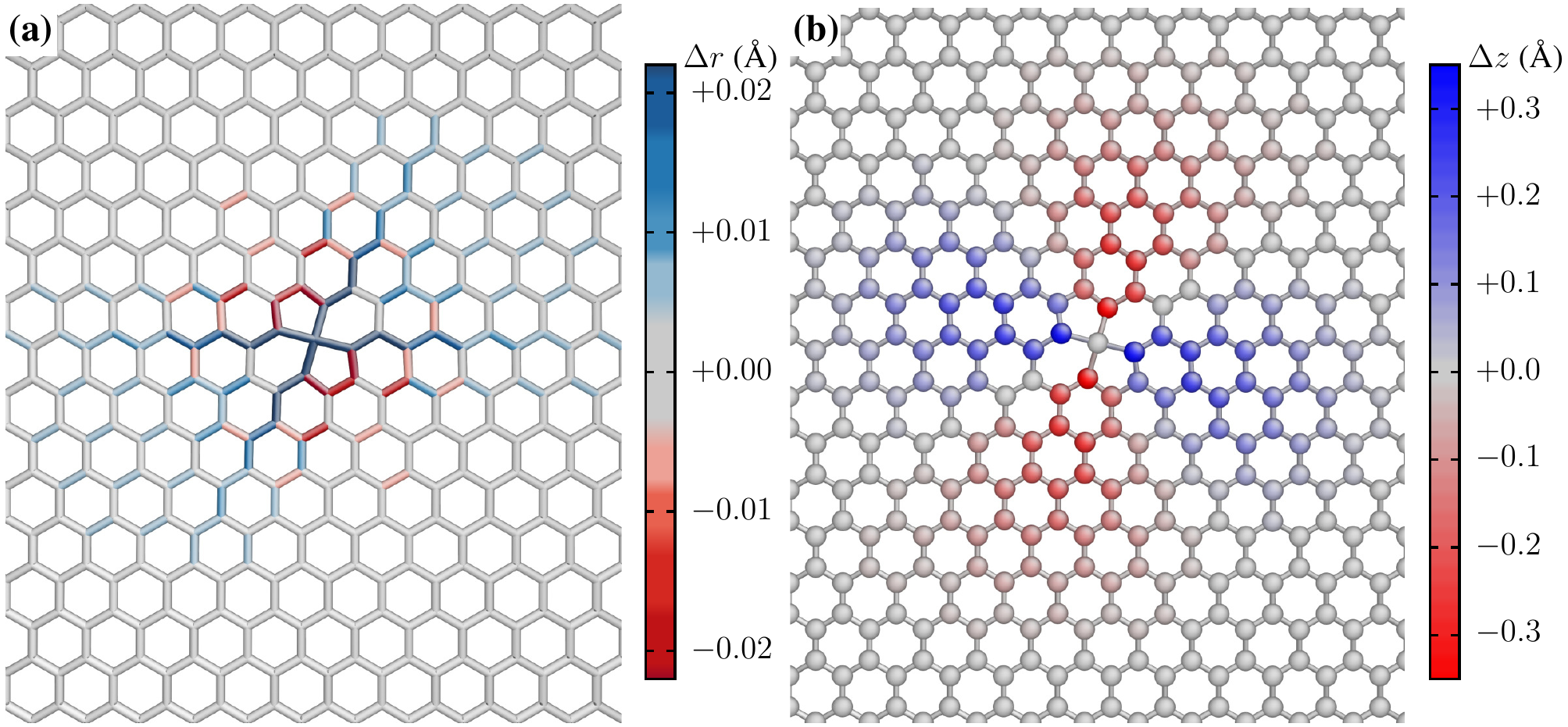}}
\caption{The complex in-plane and out-of-plane lattice relaxation at the transition state minimizes the activation barrier as shown for the 13$\times$8$(\sqrt{3},3)a_0$ supercell. (a) Elongation and contraction of bonds around the moving atom are shown in blue and red colors, respectively. (b) The atoms around the moving C-atom are symmetrically displaced up (blue) and down (red) perpendicular to the graphene plane. The displacement field is extended over 2 nm.  }
 \label{fig:displacement}
 \end{center}
 \end{figure}

\subsection{Vacancy migration and complex lattice relaxation}
Now we investigate the diffusion kinetics of an isolated V$_1(5|9)$ vacancy defect in graphene lattice, and the importance of intricate out-of-plane and in-plane lattice relaxation.  The diffusion of planer V$_1(5|9)$  defect is predominantly dictated by the motion of a single undercoordinated C-atom. However, the concurrent response of surrounding graphene lattice is quite complex and intriguing.  The activation energy $E^a_{\rm v}$ on the minimum energy path is calculated to be 0.72 eV. The corresponding first-order transition state is found to have a corrugated structure with complex in-plane and out-of-plane lattice relaxations. The migrating C-atom at the transition-state is bonded with four neighboring atoms resulting in sp$^3$-like hybridization and long-range strain field [Figure~\ref{fig:displacement}(a)].  At the transition state, the atoms around the four-fold coordinated C-atom are symmetrically displaced up and down compared to the defected graphene plane. The out-of-plane displacement is found to be as large as $\Delta z$ = $\pm$ 0.34 \AA~ for the C-atoms that are connected to the migrating atom, and the displacement field is extended over  2 nm [Figure~\ref{fig:displacement}(b)]. Such complex lattice relaxations during vacancy migration minimizes the activation barrier. To further elucidate the mechanism and the differential dependence of such complex geometrical relaxation on the activation barrier, we investigate the transition state in further detail. First, we quantify the role of lattice relaxation perpendicular to the lattice by restricting the out-of-plane relaxation while allowing the in-plane relaxation. The planer migration path cannot relieve the residual strain, and thus the corresponding TS is much higher in energy. The calculated $E^a_{\rm v}$ is found to be 89\% higher to 1.36 eV as compared to 0.72 eV, while lattice relaxation in all directions is allowed. A comparison of C--C bond distances for the transition state geometries of the planer and non-planer migration paths show that the strain relaxation in the case of unrestricted migration is easier through out-of-plane lattice relaxation.~\citep{supple}

Next, we investigate the effect of in-plane relaxation through varied supercell size while the relaxation is without any constraints. As we have discussed earlier that the long-range in-plane strain field could lead to spurious vacancy-vacancy interaction, a smaller 6$\times$4$(\sqrt{3},3)a_0$ supercell causes the $E^a_{\rm v}$ to be 23\% larger to 0.89 eV. In contrast, the $E^a_{\rm v}$ does not change for  larger 13$\times$8$(\sqrt{3},3)a_0$ supercell compared to 10$\times$5$(\sqrt{3},3)a_0$. Thus, we conclude that the lattice relaxation perpendicular to the graphene sheet and the in-plane relaxation are extremely important, and the former substantially decrease $E^a_{\rm v}$ through the corrugation of TS. 

We discuss the present results in the light of earlier theoretical and limited experimental results. In contrast to the present results, earlier first-principles calculations predicted a much higher activation barrier, in the range of 1.1--1.4 eV.~\citep{PhysRevLett.95.205501, 10.1016/j.cplett.2005.10.106, 10.1016/j.diamond.2010.06.010, C3NR06222H, 0953-8984-26-11-115303, 10.1016/j.cplett.2016.02.005}  Having discussed the importance of strain relaxations, we argue this large discrepancy to originate mainly from the improper consideration of out-of-plane and in-plane lattice relaxation during the diffusion (Figure~\ref{fig:figure1} and Figure~\ref{fig:displacement}).  In a recent calculation, the barrier was calculated to be 0.87 eV within a similar theoretical hierarchy,~\citep{10.1016/j.cplett.2016.02.005} which is still much higher than the present classical barrier of 0.72 eV. This difference is due to the spurious vacancy-vacancy interaction owing to a much smaller supercell that was considered in the earlier calculation. 

Moreover, a quantitative experimental data on the vacancy diffusion is still absent. While a V$_1(5|9)$ vacancy has been observed to move within an experimental time-scale of few hundred seconds,~\citep{10.1021/nn401113r,PhysRevB.68.144107} the prediction of diffusivity or activation barrier is difficult due to the lack of information about local temperature under TEM, and vacancy jump-rate as a function of temperature. However, we consider the vacancy diffusion on graphite surface that was studied through STM and discuss it in the context of present results on the vacancy diffusion in single-layer graphene. Using the experimental jump-rate and assuming a pre-exponent factor of 10$^{13}$ Hz, the experimental activation energy was estimated to be 0.9--1.0 eV.~\citep{10.1021/jp901578c}  Based on the present results, we propose this barrier to be treated as the upper bound for vacancy migration in single-layer graphene. We have already demonstrated that any restriction to the out-of-plane lattice relaxation during diffusion increases the barrier. Thus, as the complete out-of-plane relaxation is hindered on the basal plane of graphite, the measured activation barrier was expectedly higher than the same for single-layer graphene. 

The present results also infer that in case of a subsurface vacancy in bulk graphite, the migration barrier is expected to be similar to the planer migration barrier of 1.36 eV as the out-of-plane lattice relaxation may be completely blocked.  This claim is corroborated by the drop in vacancy kinetics that is observed through TEM for a vacancy in the middle layer of a trilayer graphene.~\citep{C4NR01918K} Similarly, a migration barrier of 1.8 $\pm$ 0.3 eV was attributed to the vacancy diffusion in irradiated graphite.~\citep{PhysRevB.47.11143} For bilayer graphene, a partial out-of-plane relaxation is still possible and the corresponding migration energies should have an intermediate value to graphite and single-layer graphene, closer to the vacancy in the basal plane of graphite.~\citep{C4NR01918K,10.1021/jp901578c,C7NR03879H,PhysRevB.82.174104}

In addition to the mechanism discussed above, a very different diffusion mechanism, migration by one zigzag lattice plane, was experimentally suggested by analyzing the TEM images.~\citep{10.1021/nn401113r}  In our calculation, such migration results with a very high migration barrier $\sim$ 3.5 eV and thus unlikely. In contrast, we argue that this new mechanism may be a combination of V$_1(5|9)$ rotation and migration that is discussed above. Further, there could be another possibility. In this experiment, the studied vacancy was separated approximately by 1 nm from a larger and more complex defect structure. Thus, the vacancy in question was under the influence of strain field generated by this complex defect and thus interacting strongly with it.  In such a situation, the vacancy may migrate very differently, and the corresponding mechanism should not be considered as the case for an isolated one. Comparing with the other 2D materials such as hexagonal boron nitride, silicene, and phosphorene,~\citep{PhysRevB.75.094104,C3NR02826G,0957-4484-26-6-065705} we conclude that the microscopic mechanism of vacancy diffusion in graphene is fundamentally different due to strong covalent bonding resulting in very different strain relaxation during migration.   

Hydrogen functionalization changes the scenario completely and the concurrent migration barrier is increased to 2.3 eV. A  closer investigation reveals that the respective transition state structure to be very different than that for the bare vacancy, where the migrating C-atom forms  $sp^3$-like bonds (Fig.~{fig:displacement}). The transition state, in this case, has an asymmetric structure and does not form such $sp^3$-like bonds resulting in higher energy TS structure.~\citep{supple} Further, while the transition state is compared with the same for the bare vacancy, the in-plane lattice relaxation is found to be much localized and the out-plane relaxation to be much smaller.~\citep{supple} All these put together increase the migration barrier for the functionalized vacancy.

\begin{figure}[t!]
 \begin{center}
\rotatebox{0}{\includegraphics[width=0.47\textwidth]{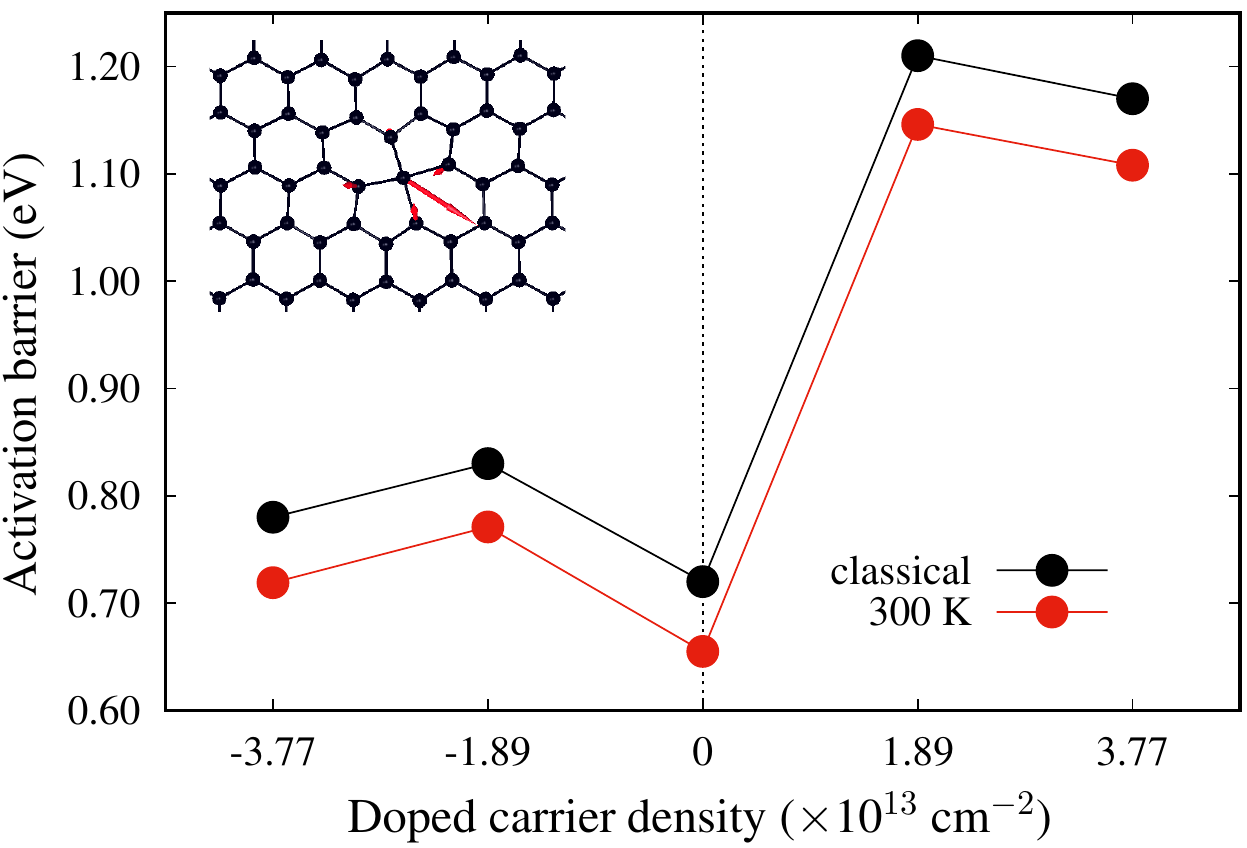}}
 \caption{Calculated activation barrier for vacancy migration in graphene lattice, and its dependence on carrier doping.  The increase in the activation barrier is much higher in case of electron doping than the corresponding hole doping. The qualitative trends with different electron and hole doping can be explained by responsible phonon modes for vacancy migration and differential charge density.  At finite temperature, the classical barrier is corrected by Wigner quantum correction and shown for 300 K. Inset shows the eigenvector for the unstable phonon mode (324$i$ cm$^{-1}$ for the neutral case) at the TS which is a superposition of both in-plane and out-of-plane lattice vibrations.}
 \label{fig:barrier}
 \end{center}
 \end{figure}

\begin{figure*}[t!]
 \begin{center}
\rotatebox{0}{\includegraphics[width=0.97\textwidth]{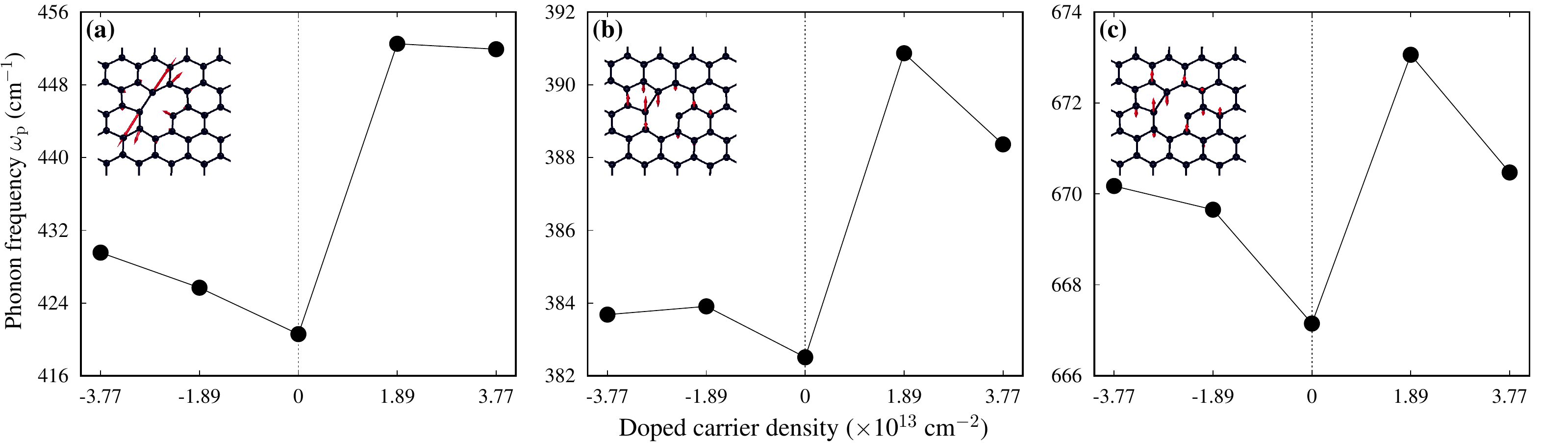}}
 \caption{Frequency evolution of the defect induced (a) in-plane and (b, c) out-of-plane phonon modes for the V$_1(5|9)$ vacancy as a function of carrier doping.  The corresponding eigenvectors are shown in the inset.The phonon modes shift to higher frequencies with carrier doping, and the effect of electron doping is stronger. Hardening of these phonon modes result in a commensurate increase in the corresponding activation barrier for a particular carrier type and concentration; and further explains the observed trends in activation barrier with varying density (Figure~\ref{fig:barrier}).}
 \label{fig:phonon}
 \end{center}
 \end{figure*}

\subsection{Effect of carrier doping}
We investigate the effect of carrier doping that can be manipulated through applied gate voltage. Here we remain within the experimentally achieved limit of carrier doping $\sim$ 10$^{13}$ cm$^{-2}$.~\citep{10.1038/nnano.2008.67} 
Though the pentagonal C--C bond of V$_1(5|9)$ is contracted or elongated upon electron or hole doping, respectively, the nature of strain field remains unaffected with extra carriers and the overall migration mechanism remains the same. However, charge doping strongly affects the activation barrier and we observe a few critical trends (Fig.~\ref{fig:barrier}). (i) The $E_{\rm v}^a$ increases with both electron and hole doping, which appears to be counter-intuitive and cannot be explained by the simple change in the local pentagonal C--C bond. (ii) While electron doping substantially affects the $E_{\rm v}^a$, hole doping has a comparatively lesser effect. (iii) The qualitative trend in $E_{\rm v}^a$ is non-monotonous and asymmetric for electron and hole doping. Electron doping substantially increases the $E_{\rm v}^a$ as much as by 68\% to 1.21 eV for 1.89$\times$10$^{13}$ cm$^{-2}$ density. Further electron doping does not alter the $E_{\rm v}^a$, which is calculated to be 1.17 eV for 3.77$\times$10$^{13}$ cm$^{-2}$ carrier density.  Similarly, the $E_{\rm v}^a$ also increases due to the hole doping, however, the increase is only moderate in contrast to the electron doping. The $E_{\rm v}^a$ increases to 0.83 eV for 1.89$\times$10$^{13}$ cm$^{-2}$ hole density, which changes to 0.78 eV while the doping is doubled. 

To explain these qualitative trends in $E_{\rm v}^a$ with carrier doping, we systematically investigate the structural, electronic and phononic properties. The local bonding picture fails to render a  comprehensive understanding of these observed trends, as the vacancy migration is a complex and collective motion of many atoms enclosing the vacancy center. Thus, we search for the low-energy phonon modes that are responsible for the vacancy migration and investigate how the frequencies of these modes change with carrier doping.  We identify three defect-induced low-energy phonon modes with frequencies ranging from 380 to 675 cm$^{-1}$ (Fig.~\ref{fig:phonon}). The eigenvector for the in-plane phonon mode explicitly indicates the characteristic signature of vacancy migration [Fig.~\ref{fig:phonon}(a)]. The undercoordinated atom moves toward the pentagonal ring and the atoms on the pentagonal edge move apart simultaneously to accommodate it and form a hexagon. The two out-of-plane phonon modes [Fig.~\ref{fig:phonon}(b) and (c)] imply the necessary lattice relaxation perpendicular to the graphene sheet during vacancy migration, which we have substantiated in the previous discussions.  

We notice the frequencies $\omega_p$ of these phonon modes increase with carrier doping (Fig.~\ref{fig:phonon}). Similar phonon hardening was observed experimentally for the C--C stretching G-mode due to both electron and hole doping realized by varied applied gate voltage.~\citep{10.1038/nnano.2008.67} Moreover, in the present case, the calculated $\omega_p$ show similar qualitative trends with varied carrier density as observed for the $E_{\rm v}^a$ (Fig.~\ref{fig:barrier}). The increase in $\omega_p$ is much higher for electron doping compared to hole doping, and a further increase in carrier concentration beyond 1.89$\times$10$^{13}$ cm$^{-2}$ does not change $\omega_p$ much. Thus, hardening of these phonon modes for both electron and hole doping leads to a commensurate increase in the calculated $E_{\rm v}^a$, and explains the qualitative trends. 

The charge density distribution of the doped carrier could also qualitatively explain the trend in $E_{\rm v}^a$ as a function of carrier concentration. Vacancy defect in graphene lattice generates semi-localized $\pi$ and $\sigma$ states, namely $V_{\pi}$ and $V_{\sigma}$, respectively.~\citep{PhysRevLett.104.096804,PhysRevB.75.125408, 10.1038/nphys2183, 10.1038/ncomms3010, PhysRevLett.117.166801} Investigating the density of states, we observe that the Dirac point shifts to $+$310 meV compared to the Fermi level for the defected graphene with a vacancy. Thus, graphene with single vacancy defect refers to an intrinsically hole doped system with one hole. Further, the spin-split $V_{\pi}$ states prevail in the vicinity of Fermi level.~\citep{PhysRevB.75.125408, PhysRevLett.117.166801}

\begin{figure}[t!]
 \begin{center}
\rotatebox{0}{\includegraphics[width=0.47\textwidth]{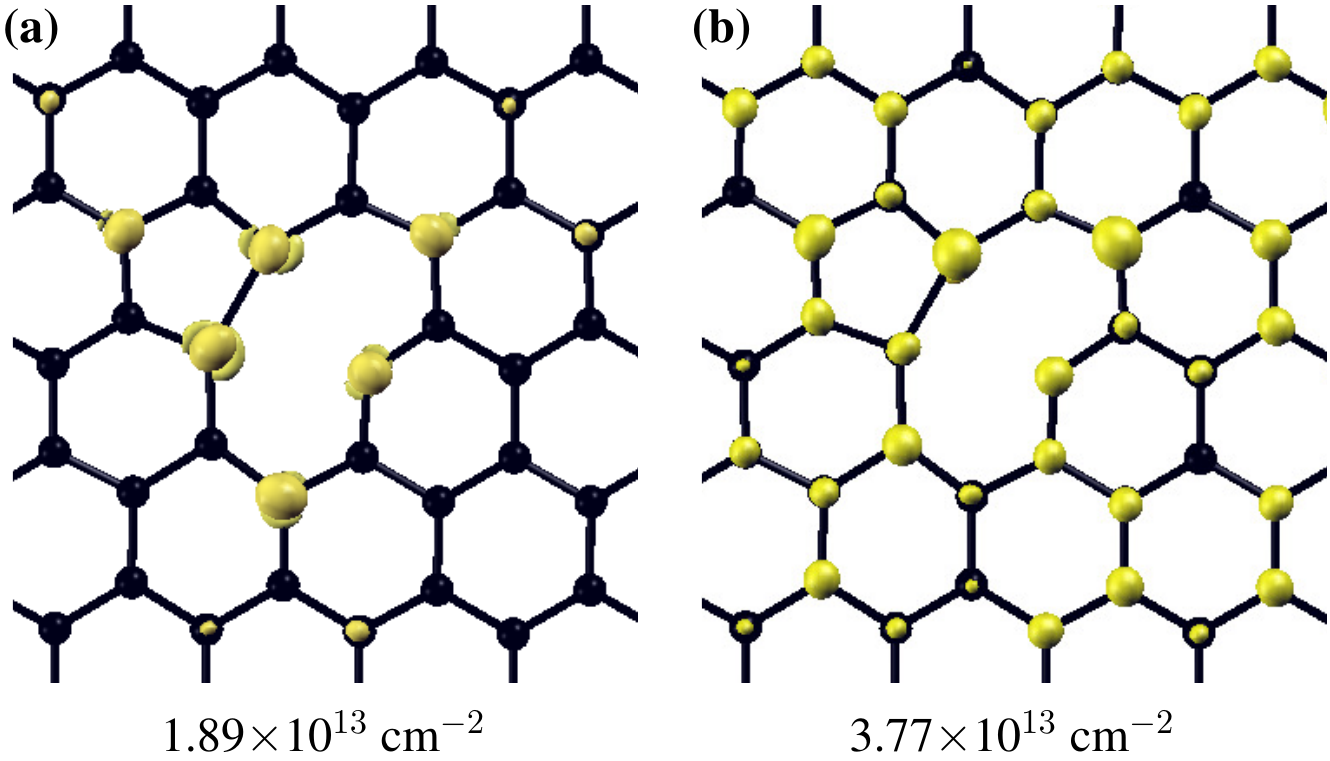}}
 \caption{The differential charge density $\Delta \rho(r)$ calculated between the neutral and the charge doped graphene, and shown for different electron doping. The doped electron is semi-localized at the V$_1(5|9)$ vacancy till 1.89$\times$10$^{13}$ cm$^{-2}$, and beyond which the charge is mostly distributed over the bulk $\pi$-state. This picture qualitatively explains the trends in phonon frequency with increasing electron doping and concurrent variation in activation barrier.  }
 \label{fig:charge}
 \end{center}
 \end{figure}

In regard to this, we calculate the differential charge density $\Delta \rho(r)$ between the neutral and charge doped graphene with a single vacancy defect.  The $\Delta \rho(r)$ determines the distribution of doped carriers, which is shown for electron doping in Fig.~\ref{fig:charge}.  While the lattice is doped with an electron, which corresponds to 1.89$\times$10$^{13}$ cm$^{-2}$ density [Fig.~\ref{fig:charge}(a)], the doped electron is semi-localized at the V$_1(5|9)$ vacancy.  Moreover, the defect induced $V_{\pi}$ state is populated, and the Fermi level coincides with the Dirac point. Thus, the semi-localized doped electron charge density distributed around the vacancy center in $V_{\pi}$ state affects the low-energy phonon modes (Fig.~\ref{fig:phonon}) that are responsible for vacancy diffusion. However, upon further increase in the electron doping beyond 1.89$\times$10$^{13}$ cm$^{-2}$, the charge is mostly distributed over the bulk $\pi$-state [Fig.~\ref{fig:charge}(b)], and thus does not alter the responsible phonon modes further. This picture qualitatively explains the trends in $E_{\rm v}^a$ with increasing electron density (Fig.~\ref{fig:barrier}).  Likewise, the trend in $E_{\rm v}^a$ with increasing hole doping could be qualitatively explained.

\subsection{Quantum correction to the activation barrier and diffusivity}
\begin{figure}[t!]
\begin{center}
\rotatebox{0}{\includegraphics[width=0.45\textwidth]{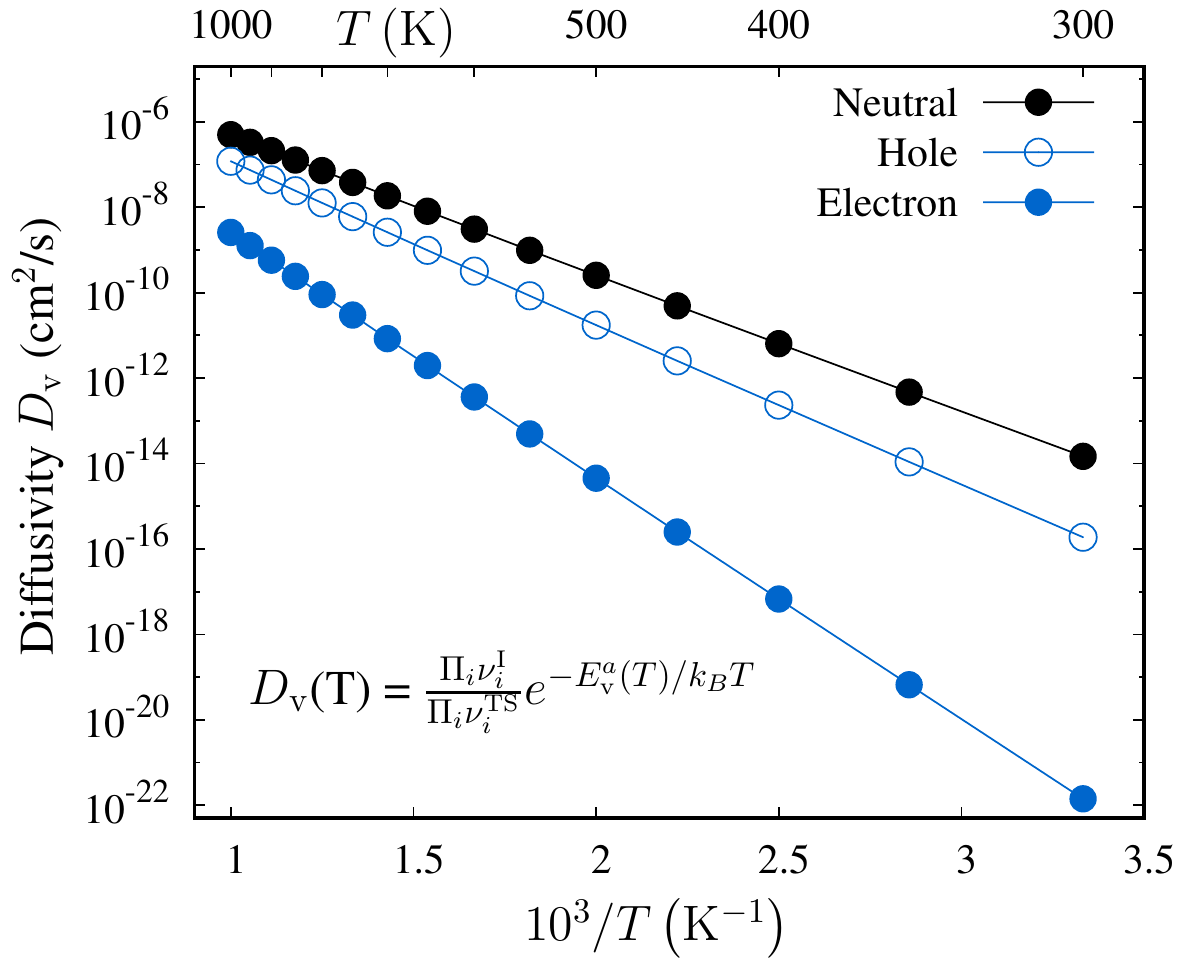}}
\caption{Diffusivity $D_{\rm v}$ calculated with Wigner corrected activation barrier $E_{\rm v}^a$(T) for neutral and carrier doped cases with 1.89$\times$10$^{13}$ cm$^{-2}$ density. The diffusivity is substantially lower for both electron and hole doping.}
\label{fig:diffusivity}
\end{center}
\end{figure}

The jump rate for vacancy diffusion in the classical transition state theory, where the vibrational modes are calculated within the harmonic approximation, is written as,~\citep{VINEYARD1957121}
\[
\Gamma_{\rm c}^{\rm hTST} = \frac{\Pi_i\nu_i^{\rm I}}{\Pi_i\nu_i^{\rm TS}}e^{-E_{\rm v}^a/k_BT}, 
\]
where $\nu_i^{\rm I}$ and $\nu_i^{\rm TS}$ are the frequencies of harmonic vibrational modes at the initial and saddle point, respectively. While the calculated activation barrier is valid at the high-temperature limit, the zero-point energy correction to the classical barrier should be invoked at very low temperatures, which can be written as, 
\[
\delta E_{\rm v, zpe}^a = \sum_i \frac{h\nu^{\rm TS}}{2} - \sum_i \frac{h\nu^{\rm I}}{2}.
\]
At any intermediate temperature, the Wigner correction and quantum tunneling should be considered. Below a critical temperature, $T^* = \hbar \nu^*/2\pi k_B$, the quantum tunneling becomes important. Here, $\nu^*$ is the magnitude of the imaginary frequency for the unstable phonon mode at the TS. For the neutral case,  $\nu^*$ is found to be 324 cm$^{-1}$, which correspond to $T^* \sim$ 75 K. 
The calculated $T^*$ monotonically changes with carrier doping and ranges between 70-86 K (Supplementary Information).~\citep{supple} Above this temperature, the Wigner correction is sufficient and the effective barrier converges to the classical value at a very high temperature. 
The Wigner correction is given by,~\citep{TF9383400029}
\[
\delta E_{\rm v, wig}^a = -k_BT \ln \left[ \frac{\Pi_i \sinh(x_i^{\rm I})/x_i^{\rm I}}{\Pi_i \sinh(x_i^{\rm TS})/x_i^{\rm TS}} \right],
\]
where $x_i = h\nu_i/2k_BT$ denotes the ratio of zero-point energy to the thermal energy for each vibrational mode. Thus,  the corrected activation barrier $E_{\rm v, corr}^a(T) = E_{\rm v}^a + \delta E_{\rm v, wig}^a(T)$,  changes with temperature (Supplemental Material),~\citep{supple} while the Wigner correction is incorporated. At room temperature 300 K, the $\delta E_{\rm v, wig}^a$ is estimated to be about 70 meV.  For example, the activation barrier for neutral vacancy decreases to 0.65 eV at 300 K compared to the classical barrier of 0.72 eV.
Using $E_{\rm v, corr}^a$ we estimate the diffusivity $D_{\rm v}$ for temperatures $T >$ 300 K (Figure~\ref{fig:diffusivity}), at which the quantum tunneling is not important. In two-dimension, $D_{\rm v}$ is given as, $D_{\rm v} = \frac{1}{4}a'^2\Gamma$, where $a'$ is the jump distance 1.84 \AA~ and $\Gamma$ is the jump rate. Due to the quantum Wigner correction, the $D_{\rm v}$ becomes one order of magnitude higher to 1.47$\times$10$^{-14}$ cm$^2$/s at 300 K than the classical diffusivity of 1.25$\times$10$^{-15}$ cm$^2$/s (Supplemental Material).~\citep{supple} 

Carrier doping considerably slows down the vacancy diffusion through a significant increase in the corresponding activation barrier (Fig.~\ref{fig:diffusivity}). The corresponding $D_{\rm v}$ decreases by a factor of $\exp(\Delta E_{\rm v}^a/k_BT)$ with $\Delta E_{\rm v}^a$ is the increase in activation barrier under carrier doping. For 1.89$\times$10$^{13}$ cm$^{-2}$ electron doping, the diffusivity is 10$^8$ times smaller than the undoped case at 300 K (Figure~\ref{fig:diffusivity}). In contrast, hole doping lowers the calculated $D_{\rm v}$ by two-orders of magnitude at the  same density and temperature. Thus, while $D_{\rm v}$ decreases for both electron and hole doping, the dramatic drop in diffusivity under electron doping will effectively freeze the vacancy and restrict diffusion.

\section{Summary}
We investigate the dynamic Jahn-Teller distortion of vacancy defect and concomitant diffusion mechanism from the first-principles calculations. The discrepancy between the present results on the dynamic Jahn-Teller distortion and the experimental observations is explained by the possible defect functionalization and charge transfer between the graphene sheet and the dielectric TEM grid.  These can hinder the observation of dynamic Jahn-Teller distortion within the experimental time-scale. We illustrate that the lattice relaxation perpendicular to the graphene sheet, as well as the in-plane relaxation, are remarkably important to predict the diffusion mechanism accurately. While the Wigner correction to the classical barrier is important to consider, the quantum tunneling can be ignored above 100 K. Influence of the applied gate voltage on the vacancy diffusion has been systematically investigated through carrier doping. The substantial increase in the activation barrier under different electron and hole doping is explained through the hardening of low-energy phonon modes that are responsible for vacancy diffusion. The qualitative trends in the activation barrier are further explained by the differential charge distribution of the doped carrier. The substantial decrease in diffusivity under carrier doping will necessarily pin the vacancy, and any degradation that is mediated by vacancy diffusion will be severely slowed down under applied gate voltage.

Experimental investigation of vacancy diffusion in two-dimension is a challenging task and the consistent interpretation of the results are very complicated. Unintentional defect functionalization and substrate effects will alter lattice relaxation and may introduce charge doping, which will essentially play a significant role. Thus, at present, the experimental attempt to quantify vacancy diffusion in graphene remains scarce. We hope the present study will motivate further experimental efforts and help interpret the results.

\begin{acknowledgements}
M. K. acknowledges the funding from the Department of Science and Technology, Government of India under Ramanujan Fellowship, and Nano Mission project SR/NM/TP-13/2016. The supercomputing facilities at the Centre for Development of Advanced Computing, Pune; Inter University Accelerator Centre, Delhi; and at the Center for Computational Materials Science, Institute of Materials Research, Tohoku University are gratefully acknowledged. 
\end{acknowledgements}


%
\end{document}